\documentclass[twocolumn,twoside,slac_two]{revtex4}
\usepackage{graphicx}
\usepackage{fancyhdr}
\begin{document}

%Title of paper
\title{}

% Repeat the \author .. \affiliation  etc. as needed
%
% \affiliation command applies to all authors since the last
% \affiliation command. The \affiliation command should follow the
% other information

\title{Triply Heavy Baryons}% Force line breaks with \\

\author{M.A. Gomshi Nobary}
\affiliation{Department of Physics, Faculty of Science, Razi
University, Kermanshah, Iran.}
\author{R. Sepahvand}
\affiliation{Department of Physics, Faculty of Science, Lorestan
University, Khoramabad, Iran.}

\begin{abstract}

The triply heavy baryons are very different in their mass. They
are essentially $\Omega_{ccc}$, $\Omega_{ccb}$, $\Omega_{cbb}$ and
$\Omega_{bbb}$ baryons which may be produced in a $c$ or a $b$
quark fragmentation. Here we briefly review the direct
fragmentation production of these states and choose $\Omega_{ccc}$
and $\Omega_{bbb}$ baryons as prototype of them to consider their
production at the hadron colliders with different $\sqrt{s}$. It
becomes clear that their production cross sections fall within a
rather diverse range according to $\sqrt{s}$, minimum transverse
momentum and rapidity. We present and compare transverse momentum
distributions of the differential cross sections, $p_T^{\rm min}$
distributions of total cross sections and the integrated total
cross sections for all triply heavy baryons at CERN LHC and simlar
quantities for $\Omega_{ccc}$ and $\Omega_{bbb}$ at RHIC, Tevatron
Run II and the CERN LHC. While some of them possess considerable
event rates even at RHIC proton-proton collider, others need much
more energetic hadron collider with appropriate kinematical cuts
to be produced.
\end{abstract}

%\maketitle must follow title, authors, abstract
\maketitle

\thispagestyle{fancy}

% body of paper here - Use proper section commands
% References should be done using the \cite, \ref, and \label commands
% Put \label in argument of \section for cross-referencing
%\section{\label{}}

\section{INTRODUCTION}

Heavy hadrons have always been interesting. Recently significant
progress have been made in understanding their production and
decay and wherever light quarks are absent, they are nicely
treated within the framework of effective field theory and
perturbative QCD. Much insights have been revealed about states
such as $B_c$, $J/\psi$ and $\Upsilon$ in due course. Recently
heavy baryons have also been focus of attention.

Heavy baryons are well distinguished according to the number of
heavy quarks involved in their formation. Essentially they are
$\Lambda_Q$'s, $\Xi_{QQ'}$'s and $\Omega_{QQ'Q''}$'s each group
with special properties [1]. Although $\Lambda_Q$'s and
$\Xi_{QQ'}$'s have been studied widely both in theory and in
experiment, the triply heavy baryons, $\Omega_{QQ'Q''}$'s have
received little attention. They are the heaviest composite states
predicted by the constituent quark model, and are the last
generation of the baryons within the standard model.

Since top quark does not materialize into hadrons [2], only $c$
and $b$ quarks take part in formation of these baryons. Therefore
the possible triply heavy baryons would be $\Omega_{ccc}$,
$\Omega_{ccb}$, $\Omega_{cbb}$ and $\Omega_{bbb}$ baryons in a $c$
or a $b$ quark fragmentation.

It is established that Heavy hadrons need to be produced at hadron
colliders with sizable cross sections and that their cross
sections is very small in $e^+ e^-$ colliders [3]. Therefore the
fragmentation production of triply heavy baryons need to be
studied in various existing and future $\bar p p$ and $pp$
colliders.

In this work we review the fragmentation production of triply
heavy baryons and have chosen mainly to emphasis on the production
of the lightest ($\Omega_{ccc}$) and the heaviest ($\Omega_{bbb}$)
as prototype them at RHIC, Tevatron Run II and the CERN LHC
colliders.

\section{FRAGMENTATION}

To evaluate the cross section of triply heavy baryons in hadron
colliders, we need their fragmentation functions. Here we have
used the established fact that the fragmentation process in
production of heavy hadrons could be understood in terms of
perturbative QCD [4]. At sufficiently large transverse momenta,
the dominant production mechanism is actually the fragmentation.
Fig. 1 shows the fragmentation of a heavy quark $Q$ into a triply
heavy baryon $B(QQ'Q'')$ in lowest order perturbation theory.

\begin{figure}
\begin{center}
\includegraphics[width=13 cm]{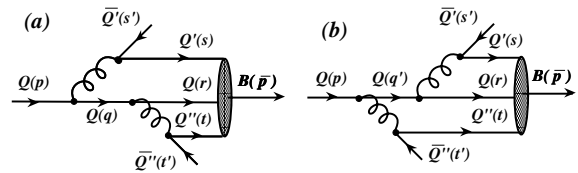}
\end{center}
\caption{ The lowest order Feynman diagrams contributing to the
fragmentation of a heavy quark ($Q$) into a triply heavy baryon
(B). The four momenta are labelled.}
\end{figure}

The fragmentation of a quark into a baryon state is described by
fragmentation function $D(z, \mu)$, where $z$ is the longitudinal
momentum fraction of the baryon state and $\mu$ is the
fragmentation scale.  The fragmentation function for the
production of an S-wave triply heavy baryon $B$ in the
fragmentation of a quark $Q$ may be put in the following form [5]

\begin{widetext}
\begin{eqnarray}
D_Q^B(z,\mu)=32\bigl[\pi^2\alpha_s({2m_{Q'}})\alpha_s({2m_{Q''}})MC_Ff_B\bigr]^2
 \int \frac{ \frac{1}{ 2}\sum\overline \Gamma \Gamma \delta^3(\overline
{\bf p}+{\bf s'}+{\bf t'}-{\bf p})} {p_\circ \overline p_\circ
s'_\circ t'_\circ(\overline p_\circ+s'_\circ+t'_\circ-p_\circ)^2}
{\rm d^3} \overline { p}{\rm d^3 }{s'}{\rm d^3 }{ t'}.
\end{eqnarray}
\end{widetext}
To obtain the above form we have convoluted the hard scattering
amplitude for Fig. 1 with an appropriate wave function for the
baryon bound state with decay constant $f_B$. The $\alpha_s$'s are
appropriate strong coupling constant and $C_F$ is the color
factor. In our model we have considered emission of two gluons by
the heavy quark $Q$, each producing a $\overline Q Q$ pair. The
three heavy quarks thus obtained form the $\Omega_{QQ'Q''}$ bound
state leaving the heavy anti-quarks to form the final state jet
(See Fig. 1). To set up the kinematics, we have used the
fragmentation parameter $z$ defined as usual, i.e.

\begin{eqnarray}
z=\frac{(E+p_\Vert)_B} {(E+p_\Vert)_Q}.
\end{eqnarray}
In an infinite momentum frame, which we have adopted for our study
here, this reduces to the following
\begin{eqnarray}
z=\frac{E_B} {E_Q}.
\end{eqnarray}

In the above general form for the fragmentation function, three
different cases are distinguished from which others could be
obtained. They are $D_{c\rightarrow \Omega_{ccb}}(z,\mu)$,
$D_{b\rightarrow \Omega_{ccb}}(z,\mu)$ and $D_{Q\rightarrow
\Omega_{QQQ}}(z,\mu)$ fragmentation functions. Where $Q$ may
assume a $c$ or a $b$ quark. The fragmentation function for
$b\rightarrow \Omega_{cbb}$ and $c\rightarrow \Omega_{cbb}$ are
obtained with interchange of $c$ and $b$ quarks in the two earlier
cases. The details of fragmentation functions for all triply heavy
baryons have been described in [5]. For simplicity here we
consider only the case for $D_{Q\rightarrow \Omega_{QQQ}}(z,\mu)$.
It has the following form [6]

\begin{widetext}
\begin{eqnarray}
D_{Q\rightarrow
QQQ}(z,\mu_\circ)&=&{{\pi^4\alpha_s^4(2m_Q)f_B^2C_F^2}\over {108
m^2z^4(1-z)^4 f^2(z) g^6(z)}}\bigl[\xi^8 z^8+4\xi^6 z^6(83-130 z+51 z^2)\nonumber\\
&&+6\xi^4z^4(1413-3084z+3022z^2-2156z^3+821z^4)+4\xi^2
z^2(18711-51678z+69417z^2\nonumber\\&&-70308z^3
+53529z^4-25950z^5+6343z^6)+222345-740664z
+1179036z^2-1253448z^3\nonumber\\&&+90126z^4-388872z^5
+109916z^6-49912z^7+20649z^8\bigr],
\end{eqnarray}
\end{widetext}

\noindent where $\alpha _s$ is the strong interaction coupling
constant evaluated at the pairs of vertices of each gluon in the
Fig. 1, $f_B$ is the baryon decay constant defined similar to
meson decay constant $f_M$ and $C_F$ is the color factor of the
baryon state formed in the fragmentation of the heavy quark.
Moreover that here we have defined $\xi=\langle{
k}_T^2\rangle/m^2$ with $k_T$ being the transverse momentum of the
initial heavy quark and $m$ is the heavy quark mass. The two
functions $f(z)$ and $g(z)$ have the following form

\begin{eqnarray}
f(z)&=& \frac{-\langle k_T^2\rangle}{{3m^2}}+ {3\over z}+
{{4}\over{3}} \biggl[1+\frac{\langle k_T^2\rangle}{
4m^2}\biggr]{1\over{1-z}},\nonumber\\
&& g(z)=-\frac{1}{3}+f(z).
\end{eqnarray}

The function $f(z)$ is the contribution of the energy denominator
emerging from the phase space integration and the function $g(z)$
is due to the quark and gluon propagators.

The inputs for the fragmentation function (1) are the quark mass,
baryon decay constant and the color factor. We have set
$m=m_c=1.25$ GeV and $m=m_b=4.25 $ GeV. For the decay constant and
the color factor we have taken $f_B$=0.25 GeV and $C_F=7/6$ for
both cases of the $\Omega_{ccc}$ and $\Omega_{bbb}$ states. We
show the behavior of (4) for $\Omega_{ccc}$ and $\Omega_{bbb}$
fragmentation along with their evolution with different scales in
Fig 2. We have also calculated the universal fragmentation
probabilities and the average fragmentation parameter, $\langle z
\rangle$. They appear in Table I.

\begin{table}
\caption{The universal fragmentation probabilities (F.P.) and the
average fragmentation parameter $\langle z \rangle$ at
fragmentation scale for different $\Omega$ states in possible a
$c$ or a $b$ quark fragmentation.}\vskip .3cm
\begin{tabular}{|c|c|c|}\hline{ Process} & F.P.
 &  $\langle z \rangle(\mu_\circ)$\\\hline
$c\rightarrow\Omega_{ccc}$&$2.789\times 10^{-5}$&0.521\\
$c\rightarrow\Omega_{ccb}$&$2.475\times10^{-6}$&0.490\\
$b\rightarrow\Omega_{ccb}$&$2.183\times10^{-4}$&0.634\\
$b\rightarrow\Omega_{bbb}$&$6.459\times10^{-7}$&0.534\\
$b\rightarrow\Omega_{cbb}$&$5.290\times
10^{-6}$&0.562\\
$c\rightarrow\Omega_{cbb}$&$1.086\times10^{-7}$&0.482\\
\hline
\end{tabular}
\end{table}

\section{INCLUSIVE PRODUCTION }

Here we have used the well known factorization scheme to evaluate
the cross sections for triply heavy baryons. Indeed the idea is to
bring about the short distance high energy parton production and
the long distance fragmentation process. All this is possible at a
scale which is much higher than the scale at which the
fragmentation functions are calculable. Therefore the
fragmentation functions calculated at fragmentation scale are
evolved up to a scale at which the convolution of parton
distribution functions, bare cross section of the initiating heavy
quark and the fragmentation function is possible. According to
this scheme in a particular scale it is possible to write
\begin{widetext}
\begin{eqnarray}
\frac {d\sigma}{dp_T}(p p \rightarrow \Omega_{QQ'Q''}(p_T)+
X)&=&\sum_{i,j}\int dx_1 dx_2 dz f_{i/p}(x_1,\mu)f_{j/
p}(x_2,\mu)\nonumber\\ &&\times\Bigl[ \hat\sigma(ij\rightarrow
Q(p_T/z)+X,\mu) D_{ Q\rightarrow \Omega_{QQ'Q''}}(z,\mu)\Bigr].
\end{eqnarray}
\end{widetext}
Where $f_{i,j}$ are parton distribution functions with momentum
fractions of $x_1$ and $x_2$, $\hat\sigma$ is the heavy quark
production cross section and $D(z,\mu)$ represents the
fragmentation of the produced heavy quark into a triply heavy
baryon. We have employed the parameterization due to
Martin-Roberts-Stiriling (MRS) [7] for parton distribution
functions and have included the heavy quark production cross
section up to the order of $\alpha_s ^3$ [8]. We have employed
this procedure in the case of RHIC, Tevatron Run II and CERN LHC
colliders. The kinematical cuts appear in Table II.

\begin{table}
\caption{The center of momentum energy ($\sqrt{s}$) and the
acceptance cuts for the colliding facilities used in this work.
The rapidity is defined as $y=\frac{1}{2}
\log\bigl\{({E-p_L})/({E+p_L})\bigr\}$. }
\begin{center}
\begin{tabular}{|c|c|c|c|}
\hline
   &RHIC& Tevatron Run II& CERN LHC\\
\hline $\sqrt{s}$ [GeV]&$\;$200&\qquad1960&$\;$\quad14000\\
 $p_T^{\rm cut}$[GeV] & $\;\;\; 2$ & $\quad\;\;\; \quad 6$& $\qquad\; 10$\\
$\quad\;\; y\leq$ & $\;\;\; $3  & $\quad\;\;\; $\quad 1&$ \qquad\;\; 1$\\
\hline
\end{tabular}
\end{center}
\end{table}

\begin{figure}
\begin{center}
\includegraphics[width=9 cm]{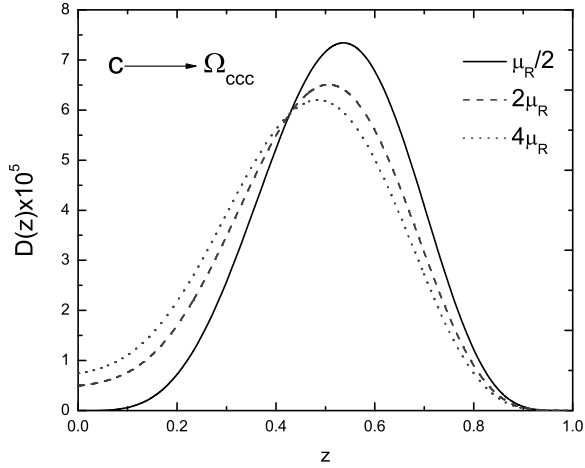}
%\hskip .25cm
\includegraphics[width=9 cm]{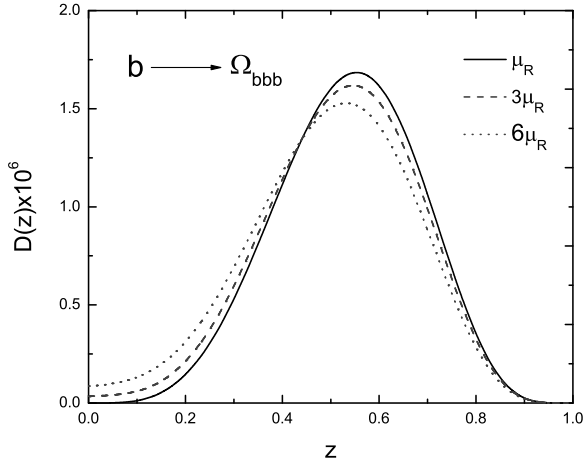}
\end{center}
\caption[Submanagers]{ The behavior of the fragmentation functions
for $\Omega_{ccc}$ and $\Omega_{bbb}$ baryons. along with their
evolutions at the scales specified.}
\end{figure}

\begin{figure}
\begin{center}
\includegraphics[width=9 cm]{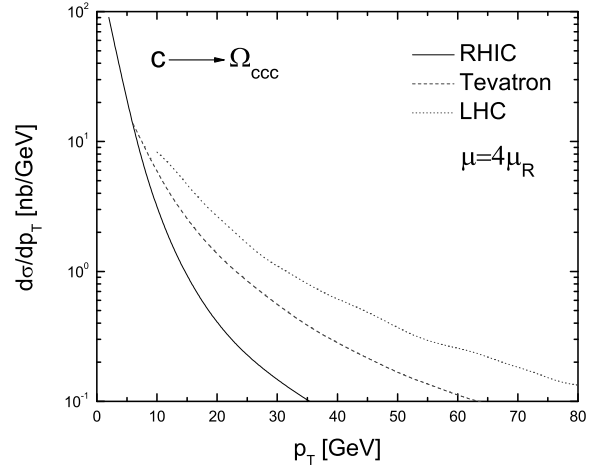}
%\hskip .25cm
\includegraphics[width=9 cm]{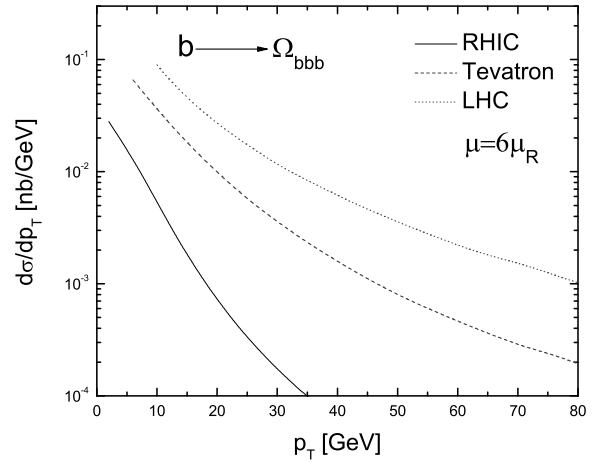}
\caption[Submanifold]{The $p_T$ distributions of the differential
cross sections in nb for $\Omega_{ccc}$ and $\Omega_{bbb}$
production at RHIC, Tevatron Run II and the CERN LHC hadron
colliders at the scale of $\mu=4\mu_R$ and $\mu=6\mu_R$
respectively. } \label{f.emb}
\end{center}
\end{figure}

\begin{figure}
\begin{center}
\includegraphics[width=9cm]{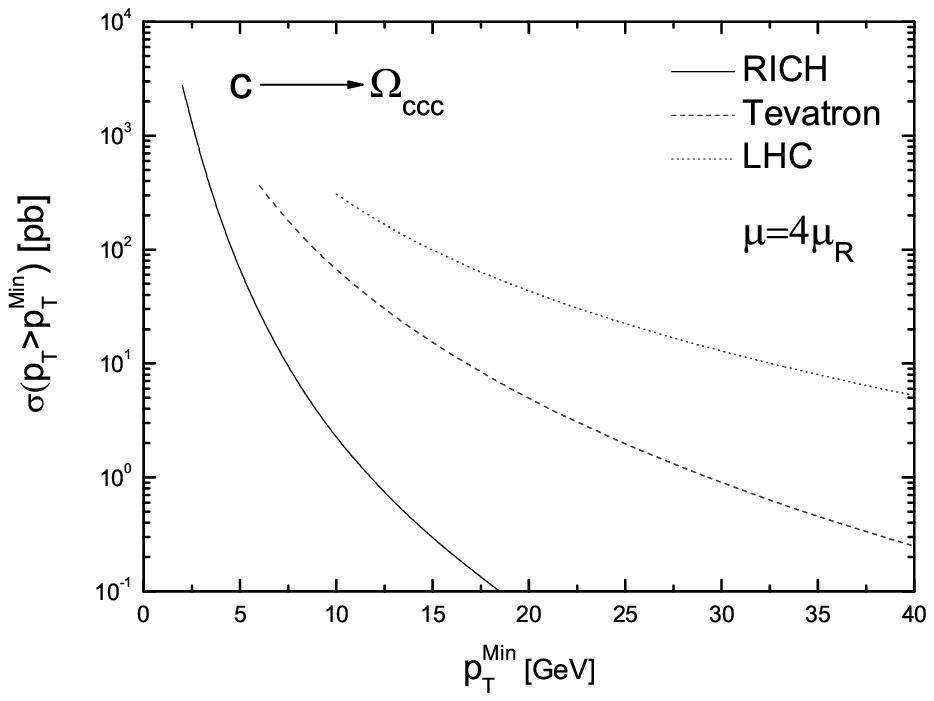}
\includegraphics[width=9cm]{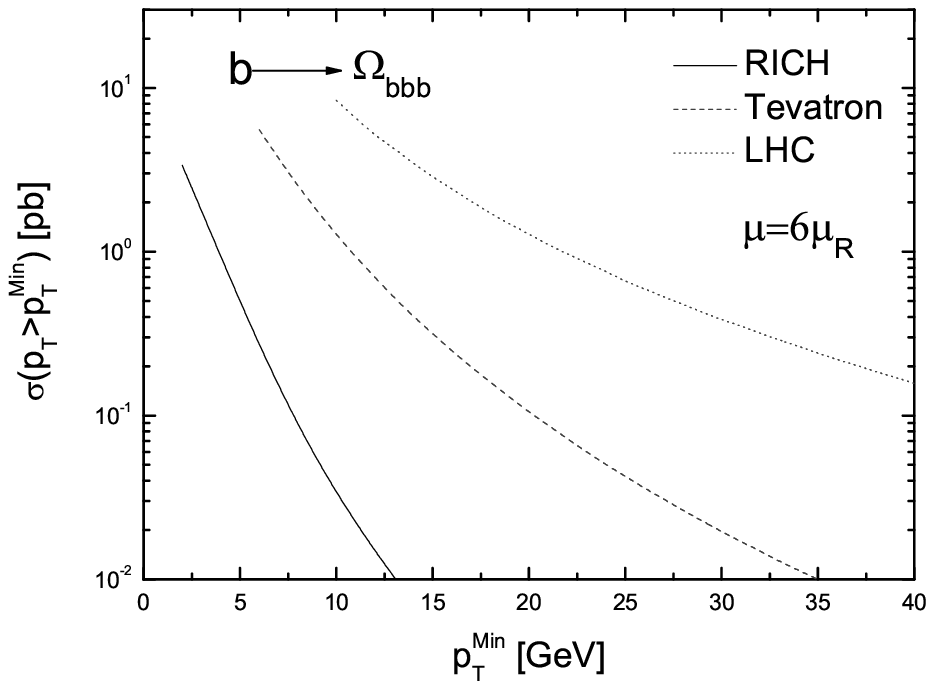}
\caption[Submanagers]{ The $p_T^{\rm min}$ distribution of the
total cross section in pb for $\Omega_{ccc}$ and $\Omega_{bbb}$
production at the RHIC, Tevatron Run II and the CERN LHC hadron
colliders at the scale $\mu=4\mu_R$ and $\mu=6\mu_R$
respectively.}
\end{center}
\end{figure}

\section{RESULTS AND DISCUSSION}

In summary we have reviewed the fragmentation functions and
production of triply heavy baryons with more emphasize on
$\Omega_{ccc}$ and $\Omega_{bbb}$ states and have evaluated their
production rates at different hadron colliders. To accomplish this
the next leading order results for parton production cross
sections are used. The fragmentation functions are calculated in
leading order perturbation. They provide reliable fragmentation
probabilities for the triply heavy baryons. In evolution of
fragmentation functions through the Altarelli-Parisi equation we
have included only the $P_{Q\rightarrow Q}$ splitting function.

The universal fragmentation probabilities and the average
fragmentation parameters at $\mu_\circ$ are shown in table I. The
probabilities at this table suggest that while some of these
states would have considerable event rates at existing colliders,
others are less probable.

Each collider with detection systems has restrictions about the
measurements of the transverse momentum and the rapidity of the
particles. The so called acceptance cuts for the colliders
considered here appear in Table II.

The behavior of our fragmentation functions along with their
evolutions at $\mu= \mu_R/2$ ($\mu_R$), $2\mu_R$ ($3\mu$ and
$4\mu_R$ ($6\mu)$ using the Altarelli-Parisi evolution equation
are shown in Fig. 2 for $\Omega_{ccc}$ and $\Omega_{bbb}$ baryons
as examples of triply heavy baryons.

Next we consider the $p_T$ distributions of the differential cross
sections at the different hadron colliders for $\Omega_{ccc}$ and
$\Omega_{bbb}$. They appear in the Figures 3. Obviously the
distributions are sensitive with respect to the scale $\mu$. The
choices of $\mu=4\mu_R$ and $\mu=6\mu_R$ respectively for
$\Omega_{ccc}$ and $\Omega_{bbb}$ are at the regions of the scale
with minimum sensitivity. Although the cross section for a given
$p_T$ differs up to three orders of magnitude from one collider to
the other,
 they are still comparable. The difference is seen
to grow with increasing $p_T$. It is more interesting in the case
of $\Omega_{ccc}$ where the distributions seem to converge at
sufficiently low $p_T$. Another important feature about these
distributions is rather high cross section of $\Omega_{ccc}$ which
is more striking in the case of RHIC where low $p_T$'s are
available. Figures 4 show the total cross sections for production
of $\Omega_{ccc}$ and $\Omega_{bbb}$ with transverse momentum
above a minimum value $p_T^{\rm min}$. Note that in the Figures 3
only the range $p_T>p_T^{\rm min}=p_T^{\rm cut}$ were considered.
It seems that the general features mentioned in the above, holds
in the case of $p_T^{\rm min}$ distributions except for that the
difference in the differential cross sections for a given $p_T$ is
not that much as in the case of  $p_T^{\rm min}$ distributions. It
is also interesting to note that the fall off of the distributions
decrease with increasing $\sqrt{s}$ and also with increasing
$p_T^{\rm min}$ or $p_T$.

We have also calculated the total cross sections. They appear in
Table III for the LHC and Table IV for RHIC, Tevatron Run II and
LHC where they are compared with each other. A short look at table
III reveals that although the total cross section for some of the
triply heavy baryons are small indeed (order of pb) and their
production needs energetic hadron colliders, some others such as
$b\rightarrow \Omega_{ccb}$ and $c\rightarrow \Omega_{ccc}$ do
possess larger cross sections of the order of nb and may easily be
produced even at the Tevatron. An interesting point in table II is
that although the total cross section for some of the particles
such as $c\rightarrow \Omega_{ccb}$ and $c\rightarrow
\Omega_{cbb}$ increase with increasing $\mu$, but this is not the
case for the rest. Our investigation shows that this depends on
the range of $\mu$ selected and also on the choice of ${p_T}^{\rm
cut}$ [9]. Note that in Table IV the cross section for
$\Omega_{bbb}$ increases with increasing $\sqrt{s}$. But this is
not the case for $\Omega_{ccc}$. Not only the order reverses in
this case but the cross section at RHIC is nearly one order of
magnitude higher.

A brief look at these tables reveal that the baryons with at least
two $c$ or at least two $b$ quarks behave similarly apart from the
magnitude of their cross sections. This is the main reason for our
choice of $\Omega_{ccc}$ and $\Omega_{bbb}$. Therefore we expect
that the $\Omega_{ccb}$ state produced in $c$ or $b$ quark
fragmentation to have similar distributions as $\Omega_{ccc}$.
Similarly the $\Omega_{cbb}$ state emerging from a $c$ or a $b$
quark will behave as $\Omega_{bbb}$. It is also interesting to
note that our chosen states $\Omega_{ccc}$ and $\Omega_{bbb}$ with
fragmentation probabilities of about $2\times 10^{-5}$ and
$6\times 10^{-7}$ are not the states with maximum and minimum
fragmentation probabilities. In other words the lightest and
heaviest in the case of triply heavy baryons does not mean the
states with maximum and minimum fragmentation probabilities.
Indeed the fragmentation probabilities for
$b\rightarrow\Omega_{ccb}$ and $c\rightarrow\Omega_{cbb}$ possess
the maximum and minimum fragmentation probabilities of $2\times
10^{-4}$ and $10^{-7}$ respectively.

\begin{table*}
\caption{Total cross section in pb for triply heavy baryons in
possible $c$ and $b$ quark fragmentation at the CERN LHC collider.
The various scales are specified.} \vskip .3cm
\begin{tabular}{|l|c|c|c|c|c|c|}\hline
 &$\mu_R/2$ &$\mu_R$&$2\mu_R$&$3\mu_R$
&$4\mu_R$&$6\mu_R$\\\hline
$c\rightarrow\Omega_{ccc}$&301.88&&306.99&&307.59&\\
$c\rightarrow\Omega_{ccb}$&26.58&&30.03&&29.88&\\
$b\rightarrow\Omega_{ccb}$&2153.08&&2155.31&&1723.80&\\
$b\rightarrow\Omega_{bbb}$&&6.34&6.38&5.77&&8.40\\
$b\rightarrow\Omega_{cbb}$&&50.30&34.77&47.78&&52.34\\
$c\rightarrow\Omega_{cbb}$&&1.14&1.38&1.47&&1.49\\\hline
\end{tabular}
\end{table*}

\begin{table}
\caption{The total cross section in pb for $ \Omega_{ccc}$ and
$\Omega_{bbb}$ baryons at different hadron colliders. The
acceptance cuts are introduced in Table II. }
\begin{center}
\begin{tabular}{|c|c|c|c|}
\hline
   &RHIC& Tevatron Run II& CERN LHC\\
\hline
 $c\rightarrow \Omega_{ccc}\;(4\mu_R)$ & 2758.26 & 382.938& 307.598\\
$b\rightarrow \Omega_{bbb}\; (6\mu_R)$ & 3.3499 & 5.91677&8.40254\\
\hline
\end{tabular}
\end{center}
\end{table}

We would like at the end discuss the uncertainties of our results.
The choice of quark masses will not only alter the fragmentation
probabilities, but also the value of $\mu$ and values of $x$ at
which the parton distribution functions are evaluated. This will
of course be reflected on the total cross sections. We have chosen
$m_c=1.25$ GeV and $m_b=4.25$ GeV which are the optimum values
reported. However the slightly higher values of $m_c=1.5$ GeV and
$m_b=4.7$ GeV are also used in the literature. Changes in quark
mass will affect the fragmentation functions. In the scheme of our
calculation, the fragmentation functions inversely depend on quark
mass squared. Therefore increase in quark mass will decrease the
probabilities. The other quantity which may depend on quark mass
is the baryon decay constant. However the later is not much clear
in the case of triply heavy baryons. Taking the explicit mass
dependence of our fragmentation functions, we have obtained 18
percent decrease in the cross sections in average, when we use the
above mentioned higher values.

There is no data on the baryon decay constant. Theoretically one
may solve the Schr\"{o}dinger like equation to obtain the wave
function at the origin for these composite particles with heavy
constituents and then relate the wave function at the origin to
the baryon decay constant. We have avoided this procedure because
of theoretical uncertainties instead have chosen $f_B=0.25$ GeV on
phenomenological grounds. The final quantity of interest is the
color factor. We have calculated this quantity using the simple
color line counting rule and have obtained $C_F=7/6$ for our
propose.

\end{document}